\newcommand{\mycomment}[1]{}
\newcommand{\cev}[1]{\reflectbox{\ensuremath{\vec{\reflectbox{\ensuremath{#1}}}}}}
\newcommand{\be}{\begin{equation}}
\newcommand{\ee}{\end{equation}}
\newcommand{\ket}[1]{\left|{#1}\right\rangle}

\documentclass[twocolumn,pra,showpacs,superscriptaddress,amssymb,amsmath,amsmath,longbibliography]{revtex4-2}
\usepackage{graphicx}
\usepackage{graphics}
\usepackage{epstopdf}
\usepackage{comment}
\usepackage{hyperref}
\usepackage{bm}
\usepackage{braket}
\usepackage{bbold}
\usepackage{inputenc}
\usepackage[normalem]{ulem}
\usepackage{appendix}

\usepackage[export]{adjustbox}

\hypersetup{pdfstartpage=1,pdfmenubar=true,pdfpagemode=None,pdftoolbar=true, urlcolor=blue,linkcolor=blue,colorlinks = true,citecolor=blue, bookmarksopen=false}

\usepackage[english]{babel}

\begin{document}

\title{Ultralong-range Rydberg molecules of Hg atoms}

\author{Agata Wojciechowska}
\email{agata.wojciechowska@fuw.edu.pl}
\affiliation{Faculty of Physics, University of Warsaw, Pasteura 5, 02-093 Warsaw, Poland}

\author{Michał Tomza}
\affiliation{Faculty of Physics, University of Warsaw, Pasteura 5, 02-093 Warsaw, Poland}

\author{Matthew T. Eiles}
\email{meiles@pks.mpg.de}
\affiliation{Max-Planck-Institut f\"ur Physik komplexer Systeme, N\"othnitzer Str. 38, 01187 Dresden, Germany}

\date{\today}

\begin{abstract}
 Ultralong-range Rydberg molecules, composed of an excited Rydberg atom and a ground-state atom, are characterized by large bond lengths, dipole moments, sensitivity to external fields, and an unusual binding mechanism based on low-energy elastic electron scattering.  Although Rydberg molecules formed between alkali atoms have received the most attention, the additional complexity found in atoms with more than a single valence electron poses new theoretical challenges as well as new possibilities for control and design of the molecular structure.  In this paper, we extend the theory of Rydberg molecules to include the additional spin coupling of the Rydberg states of a  multivalent atom. We employ this theory to describe the properties of Rydberg molecules composed of mercury atoms. We calculate the potential energy curves of both heteronuclear (Hg$^*$Rb) and homonuclear (Hg$^*$Hg) molecules. In the former case, we propose the realization of long-range spin entanglement and remote spin flip. In the latter, we show how long-lived metastable molecular states of Hg$^*$Hg exist as resonances above the dissociation threshold. 

\end{abstract}

\maketitle

\section{Introduction}
Mercury has played diverse roles in scientific discovery and application over the centuries. It served as the fluid in thermometers in its liquid form, provided the first indication of superconductivity in its solid form, and recently has found utility as a dilute ultracold gas in the context of atomic clocks and high precision measurements \cite{oskay2006single, hachisu2008trapping, gravina2022measurement,petersen2008doppler,prestage2009progress,gravina2024combpra,gravina2024comb}. 
 Hg is the heaviest element with stable isotopes to be laser cooled, and it has been magneto-optically trapped both individually \cite{stellmer2022, guo2023exploiting, mcferran2010sub, villwock2011magneto}  and simultaneously with Rb in a dual-species magneto-optical trap ~\cite{witkowski2017dual,witkowski2018photoionization}. 
 Its heavy mass and large atomic number make it a promising candidate for electron  dipole moment searches \cite{Prasannaa2015,borkowski2017optical,sunaga2019relativistic,Sunaga2019,Sahoo2018}. 
The electronic spectrum of Hg has been measured across a wide range of excited states and for a variety of term symmetries \cite{zia2003two,nadeem2009Hg_even,zia2004Hg_odd,dyubko2021Hg_F,Hg_g,clevenger1997laser}.

 Atoms excited to high principal quantum number $n$ exhibit unique properties: they are giant, long-lived, and very polarizable. 
These attributes not only broaden our fundamental understanding of atomic physics \cite{safronova2018search} but also position Rydberg atoms as promising candidates for groundbreaking applications, notably in the realm of quantum computation~\cite{saffman2010quantuminformation, morgado2021quantum, cohen2021quantum, saffman2016quantum} and quantum simulations~\cite{zoller2010RydbergSimulator, browaeys2020many}. A particularly fascinating aspect of Rydberg physics is the formation of ultralong-range Rydberg molecules in ultracold gases. 
 Formed through the coupling of a Rydberg atom and a ground-state atom via Rydberg electron scattering, these molecules were first predicted over twenty years ago~\cite{greene2000creation} and first observed in 2009~\cite{Pfaufirst2009observation}.  The investigation of Rydberg molecules has since evolved to include more sophisticated theoretical treatments, for example those including higher-order scattering effects \cite{hamilton2002shape,giannakeas2020dressed}, the coupling of electronic and nuclear spins and the fine and hyperfine structure of the constituent atoms  \cite{eiles2017J,greene2023green}, and the manipulation of these molecules via external field control \cite{bottcher2016observation,krupp2014alignment,gaj2015hybridization,hummel2018spin,hummel2019alignment}.
Rydberg molecules with many atomic constituents can also be formed, as the Rydberg electron can mediate interactions between the Rydberg atom and several ground-state atoms, as in a polyatomic molecule \cite{bendkowsky2010rydberg,eiles2016ultracold,fey2016stretching,kanungo2023measuring}, or  with hundreds or even thousands of atoms, forming a Rydberg polaron~\cite{camargo2018polaron,durst2024phenomenology,sous2020rydberg}, or with polar molecules \cite{rittenhouse2010ultracold,gonzalez2021ultralong,guttridge2023observation}.

Most of these studies of Rydberg molecules have focused on their formation from alkali-metal atoms due to the relative simplicity of manipulating and describing single-electron Rydberg series. Recently, interest in alkaline-earth-metal and other divalent atomic species has steadily grown. Rydberg molecules have been photoassociated in Sr \cite{desalvo2015Sr_pert,camargo2016lifetimes,whalen2020heteronuclear,lu2022resolving,kanungo2023measuring}, and Rydberg molecules featuring perturbed multichannel spectra in  Ca and Si have been proposed \cite{eiles2015MQDT}. 
More broadly, divalent Rydberg atoms, particularly Sr and Yb~\cite{vaillant2012yb, peper2024spectroscopy, wilson2022trapping, chen2022analyzing, nakamura2024hybrid}, have become promising candidates for quantum applications involving Rydberg atoms.
Furthermore, the promise of exciting results on quantum magnetism motivates the extension of theoretical tools to even more complex lanthanide atoms, like Er~\cite{erbium2021spectroscopy} or Ho~\cite{holmium2015measurement}. 

In this article, we propose to include Hg in the study of Rydberg molecules. 
Mercury is a divalent atom, but its spectrum is confirmed experimentally to be predominantly single-channel in character. 
It serves therefore as a good candidate for extending the spin-coupling formalism developed for alkali-metal atoms \cite{eiles2017J} to include additionally the two-electron Rydberg state, without additional complications stemming from multichannel coupling \cite{chris1996multichannel}.
We first show that the Rydberg electron of an ultralong-range Rydberg molecule can mediate an interaction between the residual valence electron of the divalent Rydberg atom and the valence electron of the ground-state atom. This occurs if the singlet-triplet splitting of the Rydberg atom becomes comparable to the hyperfine splitting of the ground-state atom, and entangles the spins of the valence electrons of the two very distant atoms. As a specific example, we study the heteronuclear Hg*Rb Rydberg molecule, but such a mechanism should be generally present in molecules involving one divalent Rydberg atom. Secondly, we show that divalent atoms add to the diversity of  Rydberg molecule structure when they are included as the ground-state atom. Through the study of the homonuclear Hg*Hg molecule, we show how Rydberg molecules can emerge as resonances above the dissociation threshold due to the interplay between positive electron-atom scattering lengths and the oscillatory structure of the potential curves. We also illustrate how Rydberg molecule spectroscopy is, in general, a useful tool to extract low-energy scattering information, since various molecular states are sensitive to small changes of the $e+$Hg scattering properties.

The structure of this article is the following. Section \ref{sec:theory} first introduces the relevant quantum numbers, focusing mainly on the heteronuclear Hg*Rb molecule. Next, it sketches the derivation of the Hamiltonian, in which we transform the Rydberg and the scattering part into a common frame in order to include the spin-orbit coupling in the electron scattering. We then show the results of the Hamiltonian diagonalization in section \ref{sec:results}. We conclude in section \ref{sec:summary}.

\begin{figure}[tb]
\begin{center}
\includegraphics[width=\columnwidth]{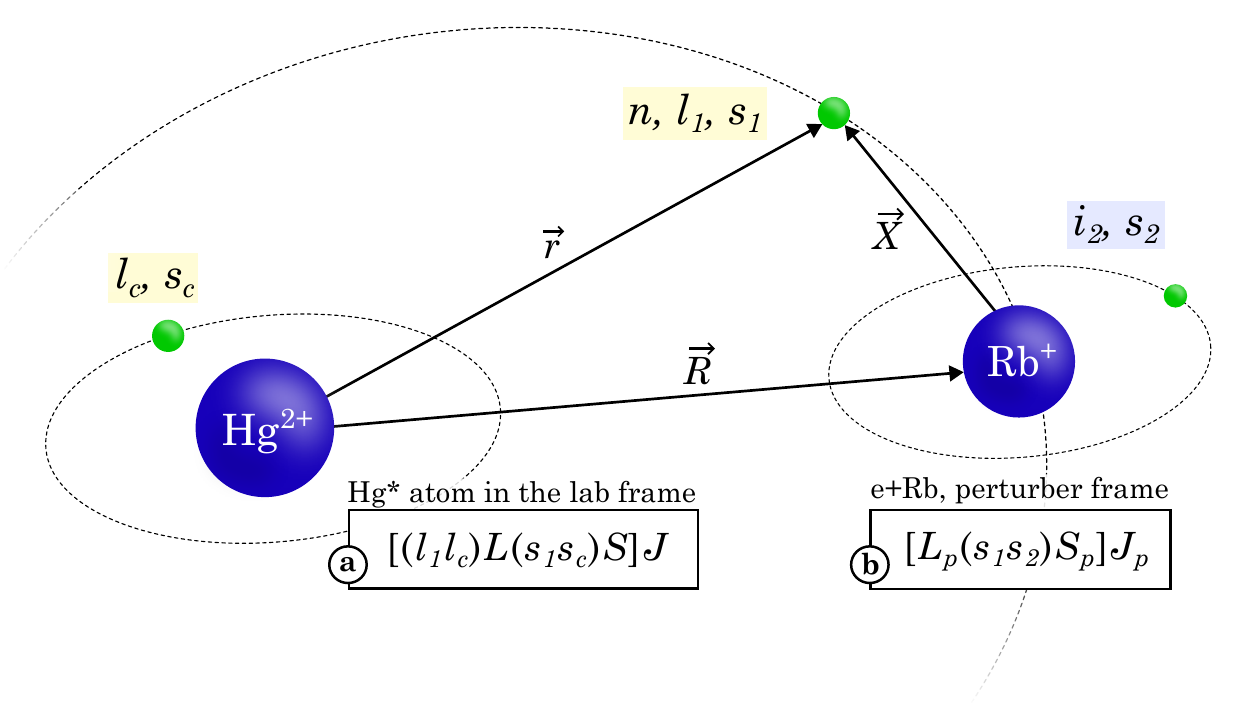}
\end{center}
\caption{Schematic drawing of the Hg*Rb Rydberg molecule. We show the ionic cores of both species and the relevant valence electrons. We introduce the quantum numbers of the Rydberg atom (in yellow) and the ground state atom (blue). The bottom box (a) contains the relevant angular momentum quantum numbers of the Hg Rydberg atom in $(LS)$ coupling and in box (b) we have the relevant quantum numbers for the Rydberg electron scattering on the perturber. We discuss them in more detail in the main text.}
\label{fig:HgRb_model}
\end{figure}

\section{Theoretical methods}\label{sec:theory}
\subsection{Quantum numbers and coupling schemes}
Let us begin by introducing the quantum numbers of the Rydberg molecule. We build a general model for the Rydberg atom in $LS$ coupling bound to an arbitrary ground-state atom ``perturber," keeping in mind that we will use it for Hg*Rb and Hg*Hg molecules. The Hg*Rb ultralong-range Rydberg molecule is sketched in Fig.~\ref{fig:HgRb_model}. The electrons in Hg form an ionic core with two outer valence shells -- one fully occupied by $5d^{10}$ electrons and one with a  remaining~$6s$ electron. The outer valence electron has an electronic spin~$s_c$ (spin projection $m_{s_{c}}$) and orbital angular momentum $l_c=0$. As we discuss later, despite this additional electronic substructure, the Rydberg states of Hg are well-described in a single channel picture in which the Rydberg electron is characterized by the principal quantum number~$n$, its orbital angular momentum $l_1$ and its projection onto the internuclear axis $m_{l_1}$, and its electronic spin quantum numbers ~$s_1$ and $m_{s_1}$. We couple the total electronic spin and total orbital angular momenta of the two valence electrons to form a state of total angular momentum $J$, $\ket{[(l_cl_1)L(s_cs_1)S]JM_J}$, as Rydberg states of Hg are well-described in $LS$ coupling.
The perturber, located a distance $R$ from the core, has a nuclear spin~$i_2$ and an electronic spin $s_2$. 
In the frame of the perturber, the Rydberg electron has an orbital angular momentum $L_p$; since the electron scatters at a very low energy, only partial waves $L_p\le 1$ need to be considered. 
The electron-atom scattering phase shifts are, in general, characterized by the total angular momentum of the negative ion complex. 
For this, we couple the orbital and electronic spin angular momenta of the Rydberg electron and the ground state atom together to form a state $|[(s_1s_2)S_pL_p]J_pM_{J_p}\rangle$. The projection of the total angular momentum onto the molecule symmetry axis, $m_{\text{TOT}}=M_J+m_{s_2}+m_{i_2}$, is conserved. 
The isotopes we use here are $^{87}$Rb, with $i_2 =3/2$, $s_2=1/2$, and a mass of 86.91 u, and the most common mercury isotope, $^{202}$Hg, with $i_2 = 0$, $s_2 = 0$, and a mass of 201.97 u.

The appropriate quantum numbers and coupling schemes used to describe the system depend on the region of space being considered.
Over most of the Rydberg volume, where the Coulomb potential dominates and the interaction of the electron with the perturber is negligible, the Hamiltonian is diagonal in the representation
\begin{equation}\label{eq:alpha}
    \alpha = \ket{n[(l_1l_c)L(s_1s_c)S]JM_J}\ket{m_{s_2}m_{i_2}},
\end{equation}
which consists of a direct product of the Rydberg electron's state and the uncoupled spins of the perturber. Near the perturber, where the Coulomb potential is locally flat and therefore irrelevant, the wave function is better described in a different basis set
\begin{equation}\label{eq:beta}
    \beta = \ket{k(L_pS_p)J_pM_{J_p}}\ket{l_cm_{l_c}s_cm_{s_c}}\ket{m_{i_2}},
\end{equation}
where $k=\sqrt{2/R-1/n^2}$ is the electron wavenumber as it scatters off of the perturber.

\subsection{The molecular Hamiltonian}

\begin{figure}[tb]
\begin{center}
\includegraphics[width=\columnwidth]{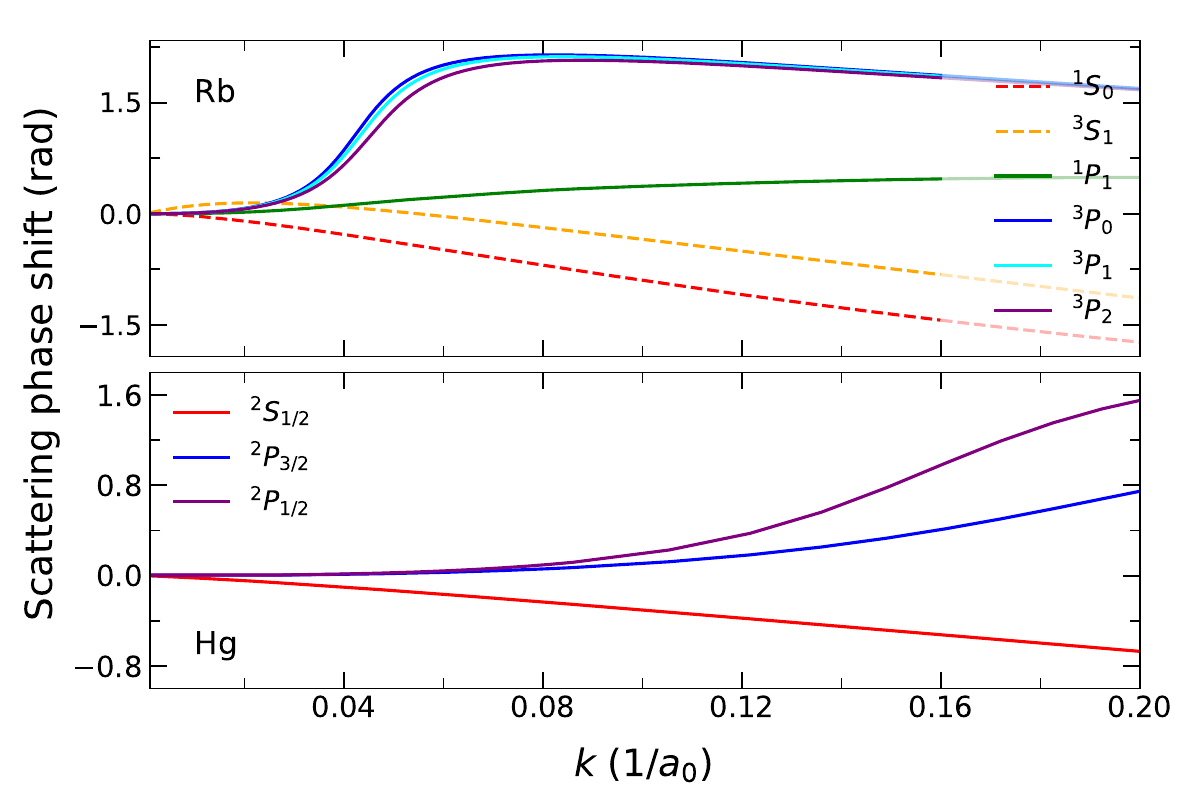}
\end{center}
\caption{Scattering phase shifts of an electron interacting with Rb (upper panel) and Hg (bottom panel). The Rb scattering data was taken from Ref.~\cite{engel2019precision}, while the Hg-$e^-$ phase shifts were extracted from Ref.~\cite{mceachran2003momentum}. The latter phase shifts were linearly extrapolated to match the zero energy scattering length. The legends give the atomic terms of the electron coupled to the perturber, i.e. $^{2S_p+1}L_{{p}_{J_p}}$ (not to be confused with the Rydberg atom term).}
\label{fig:HgRb_phase_shift}
\end{figure}

The electronic Hamiltonian reads
\begin{equation}\label{eq:H}
    \hat{H} = \hat{H}_{\text{0}} + \hat{V} + \hat{H}_{\text{HF}},
\end{equation}
where $\hat{H}_{\text{0}}$ is the Hamiltonian of the Rydberg atom, $\hat{V}$ is the Fermi pseudopotential describing the interaction between the Rydberg electron and the perturber, and $\hat{H}_{\text{HF}} = A \hat{i}_2 \cdot \hat{s}_2$ is the hyperfine interaction, with $A$ being the hyperfine constant.

The Hamiltonian of the Rydberg atom $\hat{H}_0$ is diagonal in the basis $\ket{\alpha}$ outlined in Eq.~(\ref{eq:alpha}). Its diagonal elements are
\begin{equation}
    E_{\alpha} = -\frac{1}{2(n_{\alpha}-\mu_{\alpha})^2},
\end{equation}
where $\mu_{\alpha}$ is the quantum defect of the state in basis $\ket{\alpha}$, which labels different sets of quantum numbers (in our case $S$, $L$ and $J$). In general, the quantum defects depend on the term symbol of the Rydberg atom, but in multivalent atoms they can also depend strongly on $n$ when different fragmentation channels are coupled due to level perturbations from doubly excited states.  From the Hg ground state ($5d^{10}6s^2$), a Rydberg state can be created by exciting any of the valence shell ($6s$ or $5d$) electrons. But, as shown in Ref.~\cite{baig2008high}, a $5d$ electron excited to the Rydberg state ends above the $6s$ ionization threshold. Low-lying doubly excited states where both of the $s$-shell electrons are excited have been reported, which cause small perturbations mainly to low-lying ($n=9$) $6snp(^1P_1)$ states, coupled to $5d^96s^26p(^1P_1)$, Ref.~\cite{baig1983high}. For the high $(n>20)$ principal quantum numbers most relevant for Rydberg molecules, the lack of high-lying doubly-excited states in Hg implies that its quantum defects only weakly depend on $n$. 

 The quantum defects of even-parity $S$ and $D$ states (Ref.~\cite{nadeem2009Hg_even}), odd parity $P$ states (Ref.~\cite{zia2004Hg_odd}), and $F$ (Ref.~\cite{dyubko2021Hg_F}) and even $G$ (Ref.~\cite{Hg_g}) states have all been measured in
 Hg. Table \ref{table:qd_Hg} presents the average fractional parts of these quantum defects. Due to the weak $n$-dependence of the quantum defects, the arithmetic average was taken over all of the measured $n$ levels. The large atomic number of Hg leads to a significant spin-orbit coupling.
Apart from the $S$ states and $^3P$ series, most of the quantum defects are smaller than $0.1$, which results in the accumulation of a dense spectrum of energy levels with small splittings red-detuned from the hydrogenic threshold.

The interaction $V$ is given by the Fermi pseudopotential \cite{fermi1934sopra}, generalized to higher partial waves by Omont \cite{omont1977theory}, and for singlet $S_p = 0$ and triplet $S_p = 1$ electronic states takes the form 
\begin{equation}
    \hat{V}_{S_p} = 2\pi \sum_{L_p=0}^{1} (2L_p+1) a^{2L_p+1}(S_pL_p,k) \overleftarrow{\nabla}^{L_p} \cdot 
    \delta^3(\vec{r}-\vec{R}) \overrightarrow{\nabla}^{L_p},
\end{equation}
where $a^{2L_p+1}(S_pL_p,k)$ is a (momentum) $k$ dependent scattering length for $L_p=0$ and scattering volume for $L_p=1$. They are computed from the scattering phase shifts via  $a^{2L_p+1}(S_pL_p,k) = -\frac{\tan{\delta(S_pL_p,k)}}{k^{2L_p+1}}$. There is a $p$-wave shape resonance in the low-energy electron-atom scattering cross section in the majority of atomic species used in Rydberg molecules. This makes the contribution of the $p$-wave scattering more dominant than typically expected in the low-energy region relevant to Rydberg molecule formation. In the case of Hg, the phase shifts were calculated in Ref.~\cite{mceachran2003momentum} via solution of Dirac-Fock scattering equations using static and dynamic multipole polarization potentials. These are reproduced in Fig. \ref{fig:HgRb_phase_shift}, where we used the effective range theory to interpolate between the reported scattering phase shifts and the calculated zero-energy scattering length.  The Rb phase shifts have been calculated by Fabrikant and coworkers \cite{khuskivadze2002adiabatic} and slightly modified following spectroscopic information obtained from Rydberg molecules \cite{engel2019precision}. The sign of the scattering length is important to determine whether the interaction potentials in the molecule will be attractive or repulsive. In contrast to the large negative triplet zero-energy scattering length of Rb (-15.2 $a_0$ as reported in \cite{engel2019precision}), the zero-energy scattering length of Hg is small and positive (1.87 $a_0$ as reported in \cite{mceachran2003momentum}). One more difference between the two species is that the position of the $p$-wave shape resonance in Rb is much lower in energy than it is in Hg. Thus the shape resonance in Hg only contributes at much smaller internuclear distances.

\begin{table*}[ht]
\caption{\label{table:qd_Hg}Fractional parts of Hg quantum defects for relevant $^{2S+1}L_J$ symmetries (Refs.~\cite{nadeem2009Hg_even, zia2004Hg_odd, dyubko2021Hg_F, Hg_g}), averaged over measured principal quantum numbers.}
\begin{ruledtabular}
\begin{tabular}{c|c|c|c|c|c|c|c|c|c|c|c|c|c|c|c}
\rule{0pt}{14pt} $^{2S+1}L_J$& $^0S_0$ & $^3S_1$ & $^1P_1$ & $^3P_0$ & $^3P_1$ & $^3P_2$ & $^1D_2$ & $^3D_1$ & $^3D_2$ & $^3D_3$ & $^1F_3$ & $^3F_2$ & $^3F_3$ & $^3F_4$ & $G$ \\
\hline
\rule{0pt}{9pt} $\mu_{LSJ}$ & 0.6484 & 0.6943 & 0.0503 & 0.2114 & 0.2005 & 0.0984 & 0.0777 & 0.0642 & 0.0574 & 0.0451 & 0.0291 & 0.0351 & 0.0332 & 0.0263 & 0.00655 \\
\end{tabular}
\end{ruledtabular}
\end{table*}

The mentioned scattering phase shifts on both Rb and Hg are $J_p$-dependent, as they account for the spin-orbit coupling of the electron and perturber.
Thus we will express the pseudopotential in the $\beta$ basis set of Eq.~(\ref{eq:beta}). We mediate the frame transformation following the derivation from Ref.~\cite{eiles2017J} and the detailed derivation is presented in the Appendix. It provides the matrix form of the $J_p$-dependent pseudopotential (with $J_p$-dependent scattering length and volume $a^{2L_p+1}(S_pL_pJ_p, k)$). To arrive at this form, we calculate the matrix element of the potential on the wavefunction (given in the fully uncoupled representation in the Appendix, Eq.~(\ref{eq:wf_App})). The spatial part of this wavefunction is of the form $\Phi_{nLSJ,l_1m_{l_1}}(\vec{R}) = \frac{f_{nLSJ,l_1}(R)}{R}Y_{l_1m_{l_1}}(\hat{R})$, where $\frac{f_{nLSJ,l_1}(R)}{R}$ is the radial Rydberg wavefunction which can be expressed using Whittaker functions \cite{chris1996multichannel} and spherical harmonics $Y_{l_1m_{l_1}}(\hat{R})$. In the $\alpha$ basis the pseudopotential takes the form $\hat{V} = AUA^{\dagger}$, where the matrix elements are defined as
\begin{equation}
    U_{\beta\beta'}  = \delta_{\beta\beta'} \frac{(2L_p+1)^2}{2}a^{2L_p+1}(S_pL_pJ_p, k),
\end{equation}
containing the scattering part, and
\begin{widetext}
\begin{equation}
        A_{\alpha\beta} = \sum_{M_{L_p}} b_{L_p}  Q_{L_pM_{L_p}}^{nLSJl_1m_{l_1}} C_{s_1M_{J_p}-M_{L_p}-m_{s_2}, s_2m_{s_2}}^{S_pM_{J_p}-M_{L_p}} C_{L_pM_{L_p},S_pM_{J_p}-M_{L_p}}^{J_pM_{J_p}} C_{s_1M_J-M_{L_p}-m_{s_c}, s_cm_{s_c}}^{SM_J-M_{L_p}} C_{LM_{L_p},SM_J-M_{L_p}}^{JM_J}C_{l_1M_{L_p},l_cm_{l_c}}^{LM_L},
\end{equation}
\end{widetext}
containing all Clebsh-Gordan coefficients emerging from the spin recouplings along the derivation.  It resembles a frame transformation, mediating change in the representation of the pseudopotential from basis $\beta$ to $\alpha$. We define the constant $b_{L{p}} = \sqrt{\frac{4\pi}{2L_p+1}}$ and

\begin{equation}\label{eq:Q}
    Q_{L_pm_{L_p}}^{nLSJl_1m_{l_1}}(R) = \delta_{m_{l_1}m_{L_p}} \big [ \vec{\nabla}^L( \Phi_{nLSJ,l_1 m_{l_1}}(\vec{R})) \big]^{L_p}_{m_{L_p}}.
\end{equation}
Our model can be modified to accommodate Rydberg atoms which are better described in the $JJ$ coupling scheme. The structure of the frame transformation and interaction potential would remain the same, but the Clebsch-Gordan coefficients in $A$ would change to account for the different coupling scheme. We diagonalize the Hamiltonian whose matrix elements were derived here for both hetero- and homonuclear molecules.

\begin{figure}[tb]
\centering
\includegraphics[width=\columnwidth]{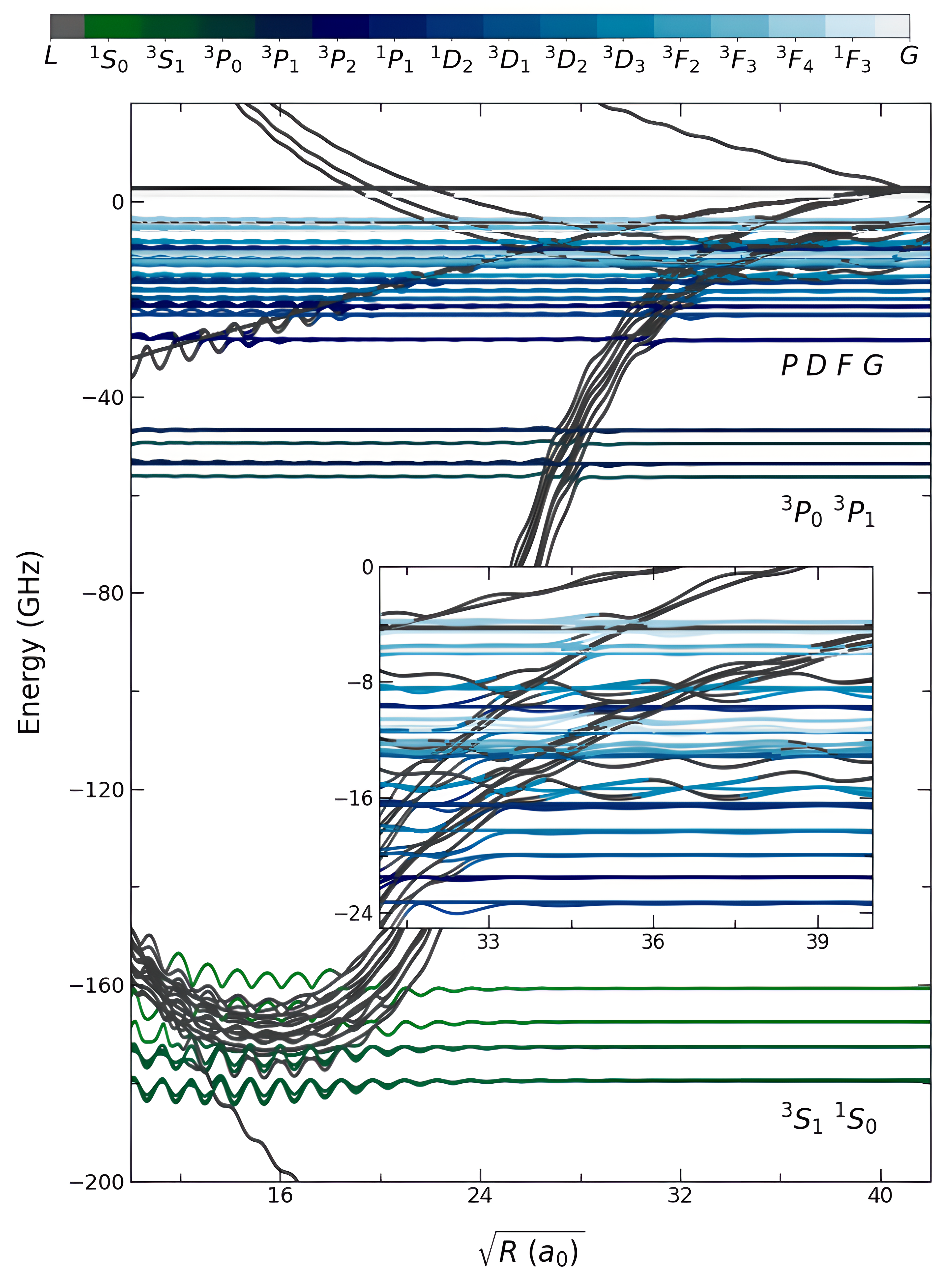}
\caption{Potential energy curves of Hg*Rb molecules with total spin projection $m_{\text{TOT}}=0$. The zero of
energy is set to the $n$ hydrogenic energy. The expected values of the Rydberg atomic term are indicated with the color code -- from greens ($S$) to light blues or grey ($F$, $G$). $L$ indicates higher angular momentum than 4. We label dissociation thresholds where space allows.}
\label{fig:HgRb_energy}
\end{figure}

\begin{figure*}[tb]
\begin{center}
\includegraphics[width=\textwidth]{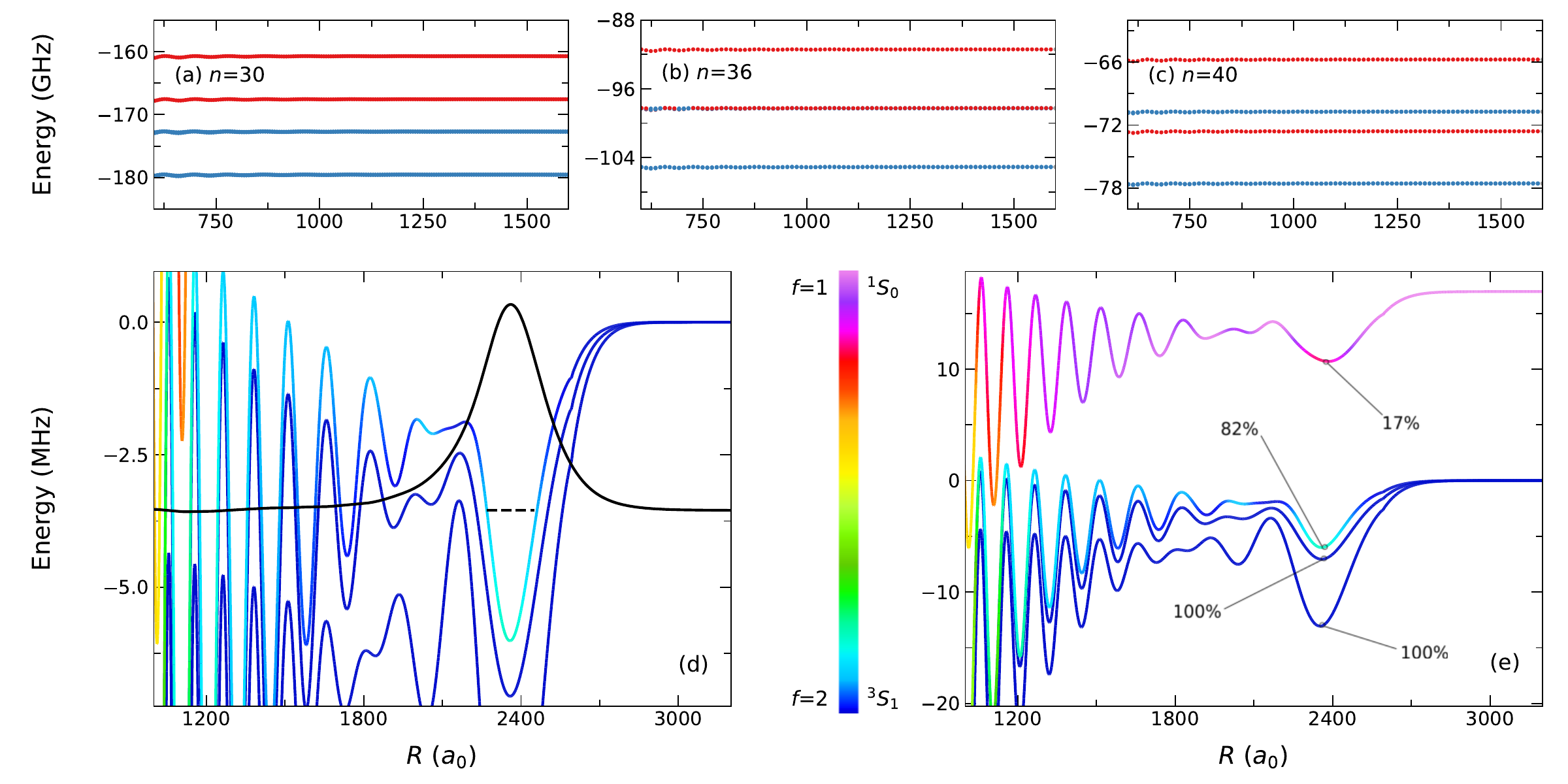}
\end{center}
\caption{Potential energy curves of the Hg*Rb molecule with $m_{\text{TOT}}=0$ and different principal quantum numbers to argue for the possible spin entanglement. The upper row (a)-(c) shows the change in the splitting of singlet (red) and triplet (blue) dominated states with $n$. For $n=36$ there in an overlap of the singlet and triplet states which we show in the bottom row. We calculate the composition of the total spin of the perturber $f$ (d) and the singlet and triplet $S$ terms (e) (also indicated with percents) to show the entanglement of spins. Finally, we show that the mixed states host localized bound states ((d), black solid wavefunction and dashed eigenenergy). In the top row the zero energy corresponds to the hydrogenic energy (with $n=30$, $36$ or $40$) and in bottom row the zero energy is shifted to match the set of triplet states.}
\label{fig:HgRb_spin_flip}
\end{figure*}

\begin{figure}[tb]
\begin{center}
\includegraphics[width=\columnwidth]{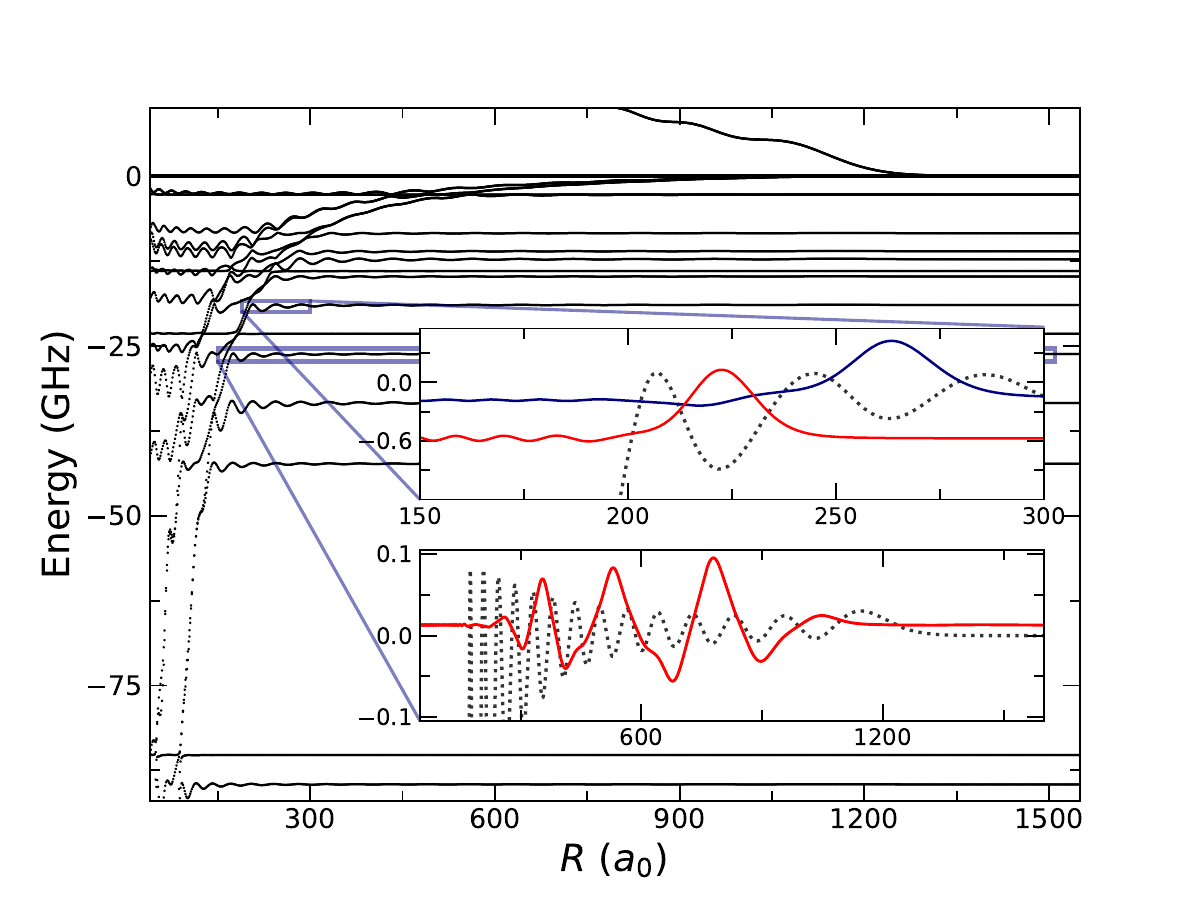}
\end{center}
\caption{Potential energy curves of the homonuclear Hg*Hg molecule with $m_{\text{TOT}}=0$, $n=25$. The insets show the two different types of vibrational states with promising applications, as discussed in the text. The upper inset shows vibrational states in the PEC of a $^3D_3$ Rydberg atom. The binding energies and stability of these molecular states are highly sensitive to the $p$-wave interaction, and thus serve as a probe of the Hg scattering phase shifts. The lower inset shows an exemplary molecular resonance in the PEC of a $^3D_1$ Rydberg atom.  This state is bound despite being energetically above the dissociation threshold. }
\label{fig:HgHg}
\end{figure}

\section{Numerical results and discussion}
\label{sec:results}

In the following calculations, we use the quantum defects from Table \ref{table:qd_Hg}. We diagonalize the Hamiltonian in a basis including four principal quantum number of interest manifolds -- one above and two below the hydrogenic manifold, because it guarantees a reasonable basis size for reliable results \cite{eiles2017J}. We express the Rydberg wavefunctions in terms of Whittaker functions for $L\leq4$ and hydrogen functions for $L>4$. 

\subsection{Hg*Rb}
We present the adiabatic potential energy curves (PECs) of the Hg*Rb molecule with $n=30$ in Fig.~\ref{fig:HgRb_energy}. As the total spin projection $m_{\text{TOT}}$ is conserved, we show results for fixed $m_{\text{TOT}}=0$. At large $R$, the interaction is zero and the different thresholds can identified by the term symbol $^{2S+1}L_J$ of the Rydberg atom and the total hyperfine spin $F$ of the ground-state atom. The clusters of PECs associated with different $L$ are labeled on the figure. We show the dominant contribution of $^{2S+1}L_J$ with a discrete color code and notice mixing, especially near the avoided crossings between PECs.

In the region under the hydrogenic threshold, exceptionally deep potential wells (in contrast to the much shallower ones in the $S$ and $P$ potential curves below) can be seen. These form because of the mixing of the many degenerate Rydberg states of high angular momentum, which creates a stronger perturbation for the Rydberg atom. Molecules bound in these potential wells are the so-called trilobite molecules \cite{greene2000creation,booth2015production}, which possess very large dipole moments due to this superposition of electronic states \cite{althon2023exploring}. 
The linear combination of electronic states forming the trilobite molecule attempts to maximize the electron density at the ground state atom.  A related state - the ``butterfly" - emerges due to the $p$-wave interaction. In this case, the electron density gradient at the perturber is maximized \cite{niederprum2016firstButterfly}. For the singlet $p$-wave scattering, where there is no shape resonance, this leads to the shallow collection of potential curves descending gradually from the hydrogenic manifold. In the case of the triplet $p$-wave, the shape resonance leads to a much larger energy shift, causing the collection of potential energy curves descending steeply from the hydrogenic manifold to appear.
Since the $S$ quantum defects are the largest, the $S$ PECs shift from the threshold and end up energetically degenerate with the butterfly wells, where we notice the spin mixing at short-range. 

Particularly intriguing phenomena unfold within those PECs, where we propose a realization of spin entanglement or remote spin-flip in the Rydberg molecule. The splitting between those PECs is determined by the interplay of the hyperfine interaction, which splits $f = 1$ and $f = 2$ states by 6.835GHz, and the Rydberg eigenenergies, which grow closer together with increasing $n$ following $n^{-3}$. At a certain $n$ value, the energy splitting between the two hyperfine levels of Rb can almost exactly match the splitting between singlet and triplet Rydberg levels. In the top row of Fig.~\ref{fig:HgRb_spin_flip}, we show these four potential energy curves for three different principal quantum numbers. The different energies match best when $n=36$, which leads to a strong mixing between the $f$ levels of the perturber and the singlet-triplet character of the Rydberg atom, giving molecular states of the form
\begin{equation}
    \alpha_1(R) \ket{^1S_0}\ket{f=1} + \alpha_2 (R)\ket{^3S_1}\ket{f=2},
\end{equation}
where $\alpha_i(R)$ indicates the amplitude of the two different spin states and can be seen in
the color code in the bottom row of Fig.~\ref{fig:HgRb_spin_flip}.
Such a state in this mixed potential curve could, for example, be excited by working with a spin-polarized gas of Rb $f=2$ ground state atoms and driving a two-photon Rydberg transition to the Rydberg triplet state. 
Such a molecular state is shown in panel (d). 
The spin of the perturber atom in a mixed state could flip, leading to an $f = 1$ state, simultaneously with a corresponding spin flip of the Rydberg atom's valence electron going from a spin triplet to a singlet state. 

A similar spin-flip was observed in Ref.~\cite{niederprum2016spin_flip} in diatomic Rb$_2$ Rydberg molecules, where mixing of spin states by the perturber atom was shown to lead to a coupling between the spin of the Rydberg electron and the spin of the ground state atom's valence electron.  Using a Hg Rydberg atom (or any two-electron atom, as long as this degeneracy condition can be met) instead has an important benefit -- the valence electron of Hg is coupled to the Rydberg electron, which in turn couples to the valence electron of the Rb ground state atom. Thus, we can produce entanglement between the spins of the valence atoms of the two atomic cores even at large ($\sim 200$nm) distances.

\subsection{Hg*Hg}
Next, using the same theoretical treatment, we obtain the energy spectrum of the homonuclear Hg*Hg Rydberg molecule shown in Fig.~\ref{fig:HgHg} for $m_{\text{TOT}}=0$.
The considered Hg ground state atom lacks nuclear spin and thus has no hyperfine structure.
The overall strength of the potential curves of Rydberg molecules scales with the principal quantum number as $n^{-a}$, where $a$ lies between $5$ and $6$. 
Thus, we present results for a smaller $n=25$ here in order to increase the amplitude of the fluctuations in the potential energy curves and compensate for the reduced strength caused by the smaller $s$-wave scattering length of Hg. 

Since the zero-energy $s$-wave scattering length is positive, the potential curves are repulsive whenever they are dominated by the $s$-wave interaction. This is typically the case at larger $R$, as seen in the bottom panel of Fig.~\ref{fig:HgHg}, showing the PEC asymptotically described by $^3D_1$ Rydberg term. In the usual case, i.e. with Rb as a perturber, one can understand the binding mechanism  as being a result of the ground-state atom finding a local maximum of electronic density and becoming trapped in it. In contrast, with an Hg perturber,  resonant states can only exist because the ground state atom becomes trapped between two repulsive barriers near a node of the electronic wave function. As long as the repulsive barriers are sufficiently large,  tunneling out of this configuration can be suppressed sufficiently to lead to metastable resonant states. Indeed, we find that these potential curves  support several weakly-confined metastable states lying above the dissociation threshold. One of these is shown in Fig.~\ref{fig:HgHg}. Such a state is trapped between regions of high electron density. 

We calculated the vibrational states using the Discrete Variable Representation (DVR). 
To confirm that such states are sufficiently long-lived to be observable, we computed lifetimes as well as resonance positions via the stabilization technique \cite{mandelshtam1993calculation}. The DVR eigenvalues are a function of the inner and outer boundaries of the DVR box; binning these eigenvalues produces Lorentzian peaks from which the resonance parameters are extracted. From this, confirmed that the lifetime of the vibrational resonances shown is on the order of that of the Rydberg atom. 

Similar metastable states have been predicted to exist in ultralong-range Rydberg trimers \cite{liu2009ultra}, but these were of a ``Borromean" nature where the well depths necessary to bind the molecule occurred only due to the presence of two ground state atoms. We expect that metastable ``above-threshold" states such as these will be interesting to probe via scattering, where they will appear as sharp resonances, or will give insight into predissociation dynamics. 
Although the state we show here appears as a type of shape resonance, lying energetically below barriers to both directions, it would be interesting to explore the possibility of long-lived resonances existing at even higher energies in the continuum. These could bear similarity with localized states to form due to quantum interference in an oscillatory potential, either as in the context of localization in quasiperiodic or disordered lattices \cite{aubry1980analyticity,anderson1958absence} or as bound states in the continuum \cite{stillinger1975bound}.

Since the $p$-wave interaction is negative, once it becomes similar in strength to the $s$-wave interaction there is a competition between attractive and repulsive forces in the binding mechanism. Close to the $p$-wave shape resonance and resulting butterfly potential energy curve, at distances around 200 a$_0$ - 500 a$_0$, this can push the potential wells below the dissociation threshold. The upper inset shows a potential curve, asociated with the Rydberg $^3D_3$ term, exhibiting this behavior and two exemplary bound states. Such states may be of particular interest to probe the collision properties of electron scattering on the ground-state Hg as their binding energies and lifetimes are highly sensitive to small changes in the scattering phase shifts. One of the shown levels (red, top inset, Fig.~\ref{fig:HgHg}) leaks out of the potential well. Using the stabilization procedure, we estimate its lifetime to be around 1200 ns.

\section{Summary and conclusions}
\label{sec:summary}

In this work, we developed the theoretical description of ultralong-range Rydberg molecule composed of a divalent single-channel Rydberg atom and a ground-state perturber. We incorporated the spin-orbit coupling of electron scattering on the ground-state atom. A physical system where this theory applies is a Rydberg molecule composed of a mercury atom, which to a good approximation does not exhibit any channel coupling in its Rydberg series. Using the Hamiltonian diagonalization method, we obtained the PECs of the Hg*Rb and Hg*Hg molecules.

The quantum defects of most symmetries of atomic Hg have small fractional parts. This, together with the hyperfine splitting of the Rb atom, produces many PECs which support bound states readily accessible via one or two-photon excitation.
Further away from the hydrogenic manifold, we encounter several PECs associated with atomic $S$ states. 
Here, we show how the spins of the valence electrons of both the Hg Rydberg atom and the Rb perturber atom can be coupled by the Rydberg electron, even across a distance exceeding $100\,$nm. This is possible due to the interplay of the splitting between the Rydberg singlet and triplet energy levels and the hyperfine splitting of the ground state atom. This can be tuned to near-degeneracy by varying the principal quantum number $n$.

Then, with the same theoretical toolbox, we calculated PECs of the homonuclear Hg*Hg molecule. Because the $s$-wave scattering length is positive, these molecules are unusual in that the long-range vibrational states are metastable resonances lying above the dissociation threshold. These could provide unusual scenarios to study molecular predissociation or even more exotic phenomena such as Anderson localization or bound states in the continuum. At smaller internuclear distances, competition between $s$- and $p$-wave scattering leads to potential curves which can support true molecular bound states. These are promising candidates for probing electron scattering properties in Hg.

Our work, on the one hand, extends the theoretical toolbox to the general form of the frame transformation for the pseudopotential and on the other hand, extends the known area of the Rydberg molecules field to a new species. Those molecules promise experimental availability, and we also plan to apply the theoretical model to even more complex atoms exhibiting multichannel structure and perturbed Rydberg series \cite{eiles2015MQDT}.

\begin{acknowledgments}
Financial support from the National Science Centre Poland (2020/38/E/ST2/00564 and 2022/45/N/ST2/03966) is gratefully acknowledged. We gratefully acknowledge Poland's high-performance computing
infrastructure PLGrid (HPC Centers: ACK Cyfronet AGH) for providing
computer facilities and support within computational grant no.
PLG/2023/016878.
\end{acknowledgments}

\bibliography{bibliografia.bib}

\clearpage
\onecolumngrid
\appendix*
\section{Derivation of the Hamiltonian}

The derivation of the Hamiltonian works both  on the tensorial form of the pseudopotential and the Rydberg wavefunction intending to express the matrix elements of the potential with the spin-orbit coupling. To do that we first need to extend the Hilbert space of the pseudopotential to the electronic spin degrees of freedom. We introduce the projection operator $\sum_{M_{S_p}}\chi^S_{M_{S_p}}(\chi^S_{M_{S_p}})^\dag$ and write 
\begin{equation}
\label{eqn:recouplingppstate}
\hat V=\mathcal{A}(S_pL_p,k) \sum_{M_{S_p}}\chi^S_{M_{S_p}}(\chi^S_{M_{S_p}})^\dag\cev\nabla {}^{L_p}\delta(\vec X)\cdot \vec\nabla^{L_p},
\end{equation}
where $\mathcal{A}(S_pL_p,k) = 2\pi(2L_p+1)a^{2L_p+1}(S_pL_p, k)$ with $a^{2L_p+1}(S_pL_p, k)$ being the energy ($k$) dependent scattering length when the angular momentum of the electron $L_p=0$ and the scattering volume -- $L_p=1$, and $\vec{X}=\vec{r}-\vec{R}$ is a vector connecting the Rydberg electron to the perturber (see Fig.~\ref{fig:HgRb_model}). Now, only by treating $\chi^S_{M_{S_p}}$ and $\vec\nabla^{L_p}$ as tensorial sets, we express the pseudopotential as a zero-rank tensor 
\begin{align}
\hat V =  \mathcal{A}(S_pL_p,k)\delta(\vec X)\sqrt{(2L_p+1)(2S_p+1)}(-1)^{-L_p-S_p} \left\{\left[\cev{\nabla}{}^{(L_p)}\otimes\vec\nabla^{(L_p)}\right]^{(0)}\otimes \left[ \chi^{(S_p)}\otimes(\chi^{(S_p)})^\dag\right]^{(0)}\right\}^{(0)}_0 .
 \end{align}
The emerging constants come from the definition of the spherical tensor components in terms of Clebsch-Gordan coefficients: $[A\otimes B]_q^k = \sum_{q_1,q_2} A_{q_1} B_{q_2} C_{1q_1,1q_2}^{kq}$. Next, we want to couple $(S_pL_p)J_p$ in one tensor product and we can do that with a simple Wigner-Racah \cite{FanoRacah,FanoMacekRMP,GreeneZareAnnRev} recoupling, in particular using the properties of the 9j-Wigner symbol. In general form, the transformation for first-rank tensors A, B, C, and D is $\{[A\otimes B]^{(0)}\otimes [C\otimes D]^{(0)}\}^{(0)}_0  = \sum_k \{[A\otimes C]^{(k)} \otimes \{[B\otimes D]^{(k)} \}^{(0)}_0 \braket{(11)k(11)k | (11)0(11)0}$.
Then also the scattering length upgrades to the spin $J_p$ dependent function $a^{2L_p+1}(S_pL_pJ_p, k)$ and in total
\begin{align}
\hat V = \delta(\vec X)\sum_{J_p}\mathcal{A}(S_pL_pJ_p,k)\sqrt{2J_p+1}(-1)^{-L_p-S_p}  \left\{\left[\cev\nabla{}^{(L_p)}\otimes\chi^{(S_p)}\right]^{(J_p)}\otimes \left[\vec\nabla^{(L_p)}\otimes (\chi ^{(S_p)})^\dag\right]^{(J_p)}\right\}_0^{(0)}.
\end{align}
Changing between tensorial and scalar operators again, we write down the scalar form of our pseudopotential with which we will be able to act on the basis set functions
 \begin{align}
 \label{eqn:approachonefinal}
   \hat V= \delta(\vec X)\sum_{J_p, M_{J_p}} \sum_{M_{L_p},M_{L_p}'} \mathcal{A}(S_pL_pJ_p,k) C_{L_pM_{L_p},S_pM_{J_p}-M_{L_p}}^{J_pM_{J_p}} C_{L_pM_{L_p}',S_pM_{J_p} - M_{L_p}'}^{J_pM_{J_p}}\cev\nabla{}^{(L_p)}_{M_{L_p}}(\chi ^{(S_p)}_{M_{J_p}-M_{L_p}})^\dag\vec\nabla^{(L_p)}_{M_{L_p}'} \chi^{(S_p)}_{M_{J_p}-M_{L_p}'}.
\end{align}
The alternative method of deriving the pseudopotential in terms of projection operators is provided in Ref.~\cite{eiles2017J}.

Now we work on the channel wavefunction to derive the matrix element of the pseudopotential. The wavefunction of the Rydberg atom (including its spatial and spin dependence) in a fully uncoupled basis reads
\begin{equation}\label{eq:wf_App}
\Psi(\vec{r}_c, \vec{r}) = \sum\limits_{\substack{m_{l_c},m_{l_1} \\ m_{s_c}, m_{s_1}}} \sum_{M_S, M_L}  Y_{l_1m_{l_1}}(\hat{r})  Y_{l_cm_{l_c}}(\hat{r}_c) \chi_{m_{s_c}}^{s_c} \chi_{m_{s_1}}^{s_1} C_{l_1m_{l_1}l_cm_{l_c}}^{LM_L} C_{s_1m_{s_1}s_cm_{s_c}}^{SM_S} C_{LM_LSM_S}^{JM_J} \frac{f_{nLSJ,l_1}(r)}{r} \frac{f_{6s}(r_c)}{r_c},
\end{equation}
where $\vec r_c$ and $\vec r$ are the position operators of the core electron and the Rydberg electron,  resepctively, $f_{nLSJ,l_1}$ is the radial wave function of the Rydberg electron and $f_{6s}$ is the radial wave function of the core electron.

Now, since the range of the interaction between the Rydberg electron and the ground state perturber is much smaller than the orbital radius of the Rydberg electron, it will be useful to employ the Taylor expansion of the wave function around the perturber. The spatial wave function of the Rydberg electron, $\Phi_{nLSJ,l_1m_{l_1}}(\vec{r}) = \frac{f_{nLSJ,l_1}(r)}{r}Y_{l_1m_{l_1}}(\hat{r})$, is expanded, yielding
\begin{equation}
    \Phi_{nLSJ,l_1m_{l_1}}(\vec{r}) \approx \Phi_{nLSJ,l_1m_{l_1}}(\vec{R}) + \vec{\nabla}[\Phi_{nLSJ,l_1m_{l_1}}(\vec{R})]\cdot \vec{X}.
\end{equation}
To first order in $|\vec X|$, the wave function reads
\begin{equation}
\begin{split}
\Psi(\vec{r_c},\vec r) = \sum\limits_{\substack{m_{l_c},m_{l_1} \\ m_{s_c}, m_{s_1}}} \sum_{M_S, M_L}  &   C_{l_1m_{l_1}l_cm_{l_c}}^{LM_L} C_{s_1m_{s_1}s_cm_{s_c}}^{SM_S} C_{LM_LSM_S}^{JM_J} \chi_{m_{s_c}}^{s_c} \chi_{m_{s_1}}^{s_1} Y_{l_cm_{l_c}}(\hat{r}_c) \frac{f_{6s}(r_c)}{r_c} \\ & \times \big[\Phi_{nLSJ,l_1m_{l_1}}(\vec{R}) + \vec{\nabla}[\Phi_{nLSJ,l_1m_{l_1}}(\vec{R})]\cdot \vec{X} \big].
\end{split}
\end{equation}
Next, we express $\hat{X}$ using spherical harmonics $Y_{L,M_L}(\hat X)$ centered about the perturber.
The wavefunction takes the form 
\begin{equation}
\Psi(\vec{r_c},\vec r) = \sum\limits_{\substack{m_{l_c} \\ m_{s_c}, m_{s_1}}} \sum_{M_S, M_L} \sum_{L_pM_{L_p}}    X^{L_p} b_{L_p} Q_{L_pM_{L_p}}^{nLSJl_1m_{l_1}}(R) \chi_{m_{s_c}}^{s_c} \chi_{m_{s_1}}^{s_1}  C_{l_1M_{L_p}l_cm_{l_c}}^{LM_L} C_{s_1m_{s_1}s_cm_{s_c}}^{SM_S} C_{LM_LSM_S}^{JM_J} Y_{l_cm_{l_c}}(\hat{r}_c) \frac{f_{6s}(r_c)}{r_c} ,
\end{equation}
where $b_{L_p} = \sqrt{\frac{4\pi}{2L_p+1}}$ and $Q_{L_pm_{L_p}}^{nLSJl_1m_{l_1}}(R) = \delta_{m_{l_1}m_{L_p}} \big [ \vec{\nabla}^L( \Phi_{nLSJ,l_1 m_{l_1}}(\vec{R})) \big]^{L_p}_{m_{L_p}}$, whose components are explicitly
\begin{align}
\label{eqn:Qfuncs}
Q_{00}^{nLSJl_10}(R)&= \frac{f_{nLSJl_1}(R)}{R}\sqrt{\frac{2l+1}{4\pi}},\\
Q_{10}^{nLSJl_10}(R) &= \sqrt{\frac{2l+1}{4\pi}}\partial_R\left(\frac{f_{nLSJl_1}(R)}{R}\right),\\
Q_{1\pm 1}^{nLSJl_1\pm 1}(R)&=\frac{f_{nLSJl_1}(R)}{R^2}\sqrt{\frac{(2l+1)(l+1)l}{8\pi}}, \ \ \ l>0.
\end{align}
Note that the cylindrical symmetry of the system imposes the constraint $m_{l_1}=M_{L_p}$. 
We next extend this wave function to include additionally the state of the perturber, $|L_qs_2\rangle$, where $s_2 = 1/2$ and $L_q = 0$ for a Rb perturber and $L_q=l_{q_1}+l_{q_2}=0$ and $s_2=s_{2,1}+s_{2,2}=0$ or $1$ for a Hg perturber. $LS$ coupling is appropriate for the ground state of both atoms. 

To ensure that we include $J_p$-dependent phase shifts, we need to couple the Rydberg electron's spin and orbital angular momenta to those of the perturber. We first couple the electronic spins together,  
\begin{equation}
    |s_1m_{s_1},s_2m_{s_2}\rangle = \sum_{S_pM_{S_p}}C_{s_1m_{s_1}s_2m_{s_2}}^{S_pM_{S_p}}|(s_1s_2)S_pM_{S_p}\rangle,
\end{equation} and then we couple $L_p$ to $S_p$, leading to
\begin{equation}
    |s_1m_{s_1},s_2m_{s_2},L_P,M_{L_P}\rangle = \sum_{S_pM_{S_p}}\sum_{J_pM_{J_p}}C_{s_1m_{s_1}s_2m_{s_2}}^{S_pM_{S_p}}C_{L_pM_{L_p}S_pM_{S_p}}^{J_pM_{J_p}}|[(s_1s_2)S_pL_pJ_pM_{J_p}\rangle.
\end{equation} Let us wrap up the transformed wavefunction
\begin{align}
\Psi(\vec{r_c},\vec r) = \sum\limits_{\substack{m_{l_c} \\ m_{s_c}, m_{s_1}}} \sum_{M_S, M_L} \sum\limits_{\substack{L_pM_{L_p} \\ S_pM_{S_p}}} \sum_{J_pM_{J_p}} &  X^{L_p} f_{L_p} Q_{L_pM_{L_p}}^{nLSJl_1m_{l_1}}(R) C_{l_1m_{l_1}l_cm_{l_c}}^{LM_L} C_{s_1m_{s_1}s_cm_{s_c}}^{SM_S} C_{LM_LSM_S}^{JM_J} C_{s_1m_{s_1}s_2m_{s_1}}^{S_pM_{S_p}} C_{L_pM_{L_p}S_pM_{S_p}}^{J_pM_{J_p}}  \\& \nonumber \times \chi_{m_{s_c}}^{s_c}\frac{f_{6s}(r_c)}{r_c} |(L_PS_P)J_PM_{J_P}\rangle.
\end{align}
We use the derived form of the wave function to calculate the matrix element of the pseudopotential. The pseudopotential takes the form of the frame transformation $V = AUA^{\dagger}$ with
\begin{equation}
    U_{\beta\beta'} = \delta_{\beta\beta'} \frac{(2L_p+1)^2}{2}a(S_pL_pJ_p, k),
\end{equation}
\begin{equation}
\begin{split}
    A_{\alpha\beta} = \sum_{M_{L_p}} & b_{L_p}  Q_{L_pM_{L_p}}^{nLSJl_1m_{l_1}} C_{s_1M_{J_p}-M_{L_p}-m_{s_2}, s_2m_{s_2}}^{S_pM_{J_p}-M_{L_p}}  C_{L_pM_{L_p},S_pM_{J_p}-M_{L_p}}^{J_pM_{J_p}}  \\ & \times C_{l_1M_{L_p},l_cm_{l_c}}^{LM_L} C_{s_1M_J-M_{L_p}-m_{s_c}, s_cm_{s_c}}^{SM_J-M_{L_p}} C_{LM_{L_p},SM_J-M_{L_p}}^{JM_J}.
\end{split}
\end{equation}

\end{document}